# Multiscale reduced-order modeling of fused filament fabricated composites

Satyajit Mojumder[a,f], Anton van Beek[b], Zahabul Islam[c], Dong Qian[d,f], Wing Kam Liu[e,f*]

[a]Theoretical and Applied Mechanics Program, Northwestern University, Evanston, IL, USA

[b]School of Mechanical and Materials Engineering, University College Dublin, Dublin, Ireland

[c]Mechanical and Manufacturing Engineering Technology, Bowling Green State University, Bowling Green, OH, USA

[d]Department of Mechanical Engineering, University of Texas at Dallas, TX, USA

[e]Department of Mechanical Engineering, Northwestern University, Evanston, IL, USA

[f]HIDENN-AI, LLC, IL, USA

**Abstract:** Defects such as voids are observed at multiple length scales of an additively manufactured composite material. Modeling such defects and their multiscale interaction is crucial for the material's performance prediction. In this work, we study as-built defects in fused filament fabricated Polycarbonate/Short Carbon Fiber (PC/SCF) composite samples. The microscale and mesoscale voids along with the mesoscale layer orientations have been studied using a mechanistic reduced-order model. Our result indicates that the microscale intrabead voids interact with the mesoscale interbead voids and significantly degrade the mechanical response of the printed composites compared to the microscale microstructure without voids. The mesoscale layer orientations also influence the stress-strain response and show better performance when the load is applied to the bead direction. The efficient reduced-order modeling approach used in this work provides a way to evaluate multiscale design aspects of additively manufactured composite materials.

## 1. Introduction

The fused filament fabrication (FFF) of composite materials has been demonstrated for different industrial applications such as aerospace, biomedical devices, and consumer good parts [1] in recent times. Defects are an inherent part of any manufacturing process and FFF is no exception[2]. These defects and microstructural features control the printed part performance and understanding their behavior is crucial for materials selection in any engineering application.

---

*Corresponding author: Wing Kam Liu (w-liu@northwestern.edu)

A very limited range of engineering thermoplastic polymers (e.g., acrylonitrile butadiene styrene (ABS), Polyamide (PA), polylactic acid (PLA), etc.) has been tested for the FFF process with fillers such as nanoparticles, nanotubes, and short carbon fiber (SCF) to enhance the mechanical properties of printed composites. The FFF process induces inherent defects such as partial neck growth interbead voids at the mesoscale level [2]. Adding fillers in the filament also introduces the intrabead voids that exist even after printing [3,4]. These microscale voids are generated due to the trapped gas and depend on the choice of processing conditions.

To predict the mechanical properties of the printed materials, researchers mostly adopted a homogenization-based multiscale analysis approach varying materials system [5], and process parameters [6–9] such as raster angle, build orientation, and infill density to study their effects. Most of these approaches consider a unit cell of the material's microstructure and used the homogenized properties in the macro scale analysis to predict the elastic properties [10–13]. Considering the detailed microstructure features such as defects at multiple length scales are still missing in the literature.

While homogenization-based multiscale analysis predicts the properties of the materials with sufficient fidelity, they are computationally expensive for modeling complex microstructure features such as defects [11]. A reduced-order model (ROM) can be an effective alternative to compute the material's response efficiently. Monaldo and Marfia [14,15] introduced the transformation field analysis-based ROM for additively manufactured microstructures. While the ROM provides the material's response faster, the accuracy depends on the type of ROM techniques used. Typical data-driven techniques such as neural networks [16,17], proper generalized decomposition, and principal component analysis [18] require an offline training dataset and the models are less predictive outside the training range. On the other hand, mechanistic reduced-order models such as self-consistent clustering analysis (SCA) [19–23] and its variant, multiresolution clustering analysis (MCA)[24] are based on the mechanistic Lipmann-Schwinger equation [25] which makes these ROM predictive for arbitrary loading conditions with a fast online computation on a pre-computed offline database. More advanced machine learning-enabled ROM methods such as HiDeNN [26,27], HiDeNN-TD [28], and C-HiDeNN-TD [29] have the potential to solve large-scale problems in computational mechanics.

As different processing conditions along with the wide choice of polymers and fillers introduce a high-dimensional materials design problem, an efficient ROM technique is demonstrated in this work to explore high-dimensional materials design space. In Section 2, the ROM development for the FFF fabricated PC/SCF composites is discussed. The effects of defects at meso- and micro-scale are discussed in section 3. Finally, in section 4, a summary is offered on the effectiveness of the ROM for the evaluation of high-dimensional materials design space for new materials design and discovery.

## 2. Method

In this work, the detailed microstructures at mesoscale (layered bead structure) and microscale (single bead) with defects for FFF printed PC/SCF composites are considered. The multiscale modeling includes an experimental microstructure characterization and reconstruction step following a ROM development for the meso- and micro-scale of the printed composite microstructures.

### 2.1. Microstructure characterization and reconstruction with defects

To understand the types of defects in different scales and their microstructure characterization purpose, PC/SCF composites with a 10% SCF volume fraction has fabricated using the FFF process with nozzle temperature 260ºC, 0.3 mm hatch spacing, and 0.2 mm layer thickness (See Fig. 1c to 1f). The dominant features of the microstructures at the mesoscale can be characterized by the shape of the beads and the printing pattern. In this work, we assume that the beads can be described as rods with an ellipsoidal cross section and infer the minor and major axis lengths from the experimental images presented in Fig. 1a. Consequently, the minor and major axis lengths can then be used in a voxel-based reconstruction framework where each bead is parametrically defined as a rod with an ellipsoidal cross-section. By iterating over the reconstruction space, we can determine which voxels are filaments and which voxels are void.

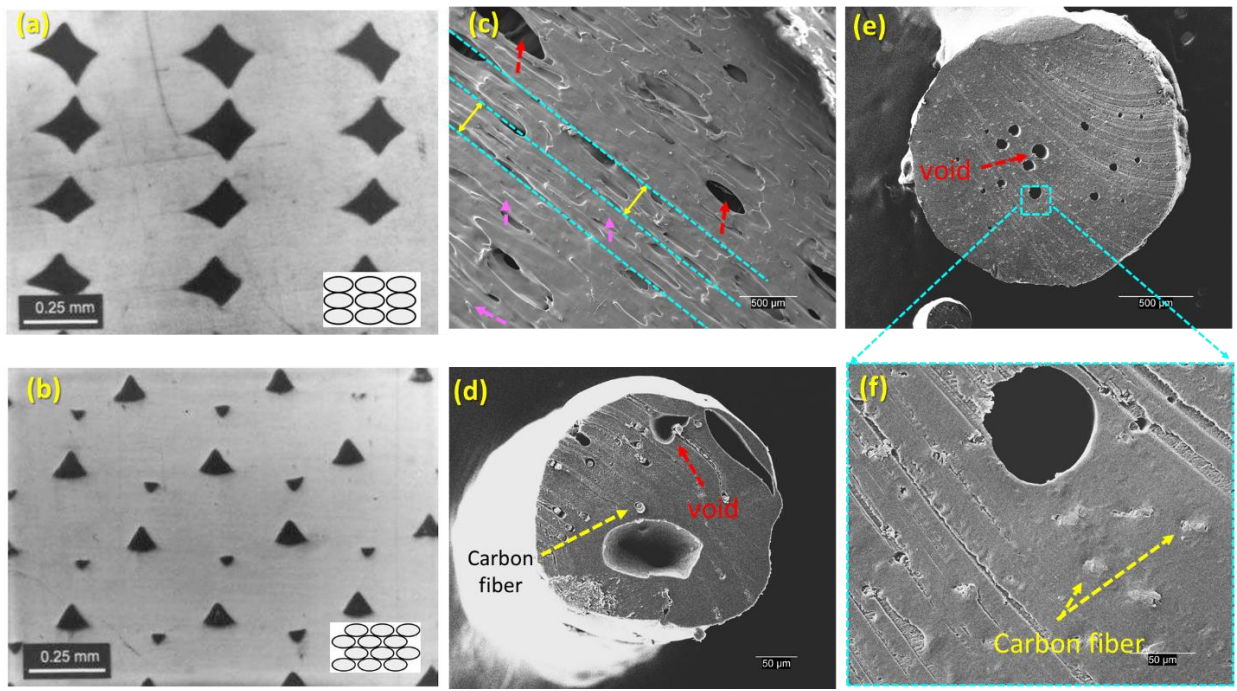

Fig. 1: Defects in the fused filament fabrication of short fiber composites. (a) rectangular printing pattern (shown in inset) produces the diamond shape void, (b) skewed printing pattern(shown in inset) produces triangular shape void [30], (c) FFF printed PC/SCF composites with 10% fiber volume fraction (d) voids in a single filament after printing, (e, f) voids in single filament before printing.

While the mesoscale reconstruction follows a periodic pattern of beads, the short fibers (and voids) in the microscale follow a random pattern. Specifically, we assume that the chopped fibers are cylindrical in shape with fixed length and aspect ratio (see Fig. 1d and 1f). Subsequently, we can then create new structures by sequentially adding cylindrical fibers to an empty structure by randomly sampling their locations and orientations from a multivariate uniform distribution. Before adding a new fiber, we verify if its sampled location and orientation will result in an intersection with fibers already present in the structure (i.e., particles get added if no intersection is observed, and otherwise a new location and orientation are sampled). This is a process known as random sequential adsorption [31] and is repeated until the volume fraction of short fibers in the reconstruction is consistent with the target structures (i.e., Fig. 1). Finally, spherical voids, with random diameters are added to the structures until the desired volume fraction of voids has been achieved. A reconstructed meso- and micro-scale microstructure is shown in Fig. 2.

## 2.2. Multiscale ROM development

Fig. 2a represents the micro-scale reconstruction of PC/SCF microstructure with 2% voids and a mesoscale bead structure with four layers printed in [0 0 0 0] angle orientation. These two scales are represented by two microstructure volume elements (MVEs) and can be considered boundary value problems (BVPs) with periodic boundary conditions. Therefore, the equilibrium BVPs on these MVEs are given by

$$\begin{cases} \boldsymbol{\nabla}^{(n)} \cdot \boldsymbol{\sigma}^{(n)} = 0, \forall \boldsymbol{X}^{(n)} \in \Omega^{(n)} \\ \boldsymbol{\varepsilon}^{(n)} = \frac{1}{2}\left(\nabla \boldsymbol{u}^{(n)} + \left(\nabla \boldsymbol{u}^{(n)}\right)^T\right), \forall \boldsymbol{X}^{(n)} \in \Omega^{(n)} \\ \boldsymbol{u}^{(n)} periodic\ on\ \partial\Omega^{(n)} \\ \boldsymbol{t}^{(n)} anti-periodic\ on\ \partial\Omega^{(n)} \end{cases} \quad (1)$$

Where (n) denotes the nth scale with n=1 as mesoscale and n=2 as microscale. This formulation is general and can be extended for any arbitrary number of scales. The coupling of the BVPs among two consecutive scales can be expressed as

$$\begin{cases} \boldsymbol{\varepsilon}^{(n-1)}(\boldsymbol{X}^{(n-1)}) = \frac{1}{|\Omega^{(n)}|} \int_{\Omega^{(n)}} \boldsymbol{\varepsilon}^{(n)}(\boldsymbol{X}^{(n)}) d\boldsymbol{X}^{(n)}, \\ \boldsymbol{\sigma}^{(n-1)}(\boldsymbol{X}^{(n-1)}) = \frac{1}{|\Omega^{(n)}|} \int_{\Omega^{(n)}} \boldsymbol{\sigma}^{(n)}(\boldsymbol{X}^{(n)}) d\boldsymbol{X}^{(n)}, \end{cases} \quad (2)$$

Where $\boldsymbol{X}^{(n-1)}$ is an arbitrary material point, and $\Omega^{(n)}$ is the nth scale MVE domain associated with $\boldsymbol{X}^{(n-1)}$. The finer scale microstructure needs to be resolved first and the homogenized response is passed as a single material point response to a higher scale. In this approach, the constitutive law for the lower length scale is provided as input and constitutive laws for the higher length scales are computed by homogenization.

To model this coupled meso- and micro-scale MVEs, a multiresolution clustering analysis (MCA) based reduced-order modeling approach is taken. Interested readers can find the details of the method in the previous work of Yu et al. [24] and Gao et al. [32]; however, we only present the key highlights of the method in this present work.

The key concept of the MCA is to decompose the high-fidelity microstructure domain into several clusters that are mechanically "similar" (having similar strain concentration or other mechanical quantities of interest). Based on this principle, the micro- and mesoscale BVPs as described in Eq. (1) are solved using the Lippmann-Schwinger equation as follows

$$\Delta \boldsymbol{\varepsilon}^{(n)}(\boldsymbol{X}) = \Delta \tilde{\boldsymbol{\varepsilon}}^{(n)} - \int_{\Omega} \boldsymbol{\Gamma}^{(n)}(\boldsymbol{X}, \boldsymbol{X}') : \left( \Delta \boldsymbol{\sigma}^{(n)}(\boldsymbol{X}') - \boldsymbol{C}^{0,(n)} : \Delta \boldsymbol{\varepsilon}^{(n)}(\boldsymbol{X}') \right) d\boldsymbol{X}', n = 1, \dots, N, \quad (3)$$

Where $\boldsymbol{\Gamma}$ is the Green's operator. The strain concentration tensors for these high-fidelity microstructures are computed using Eq. (3) for the elastic step and based on that the domains are discretized into a few clusters ($N_c^{(n)}$) for each scale n. All variables are considered uniform in a cluster and can be computed by the volume averaging technique. For any variable, the clustering variable can be approximated as

$$\square^{(n)}(X^{(n)}) = \sum_{I^{(n)}}^{N_c^{(n)}} \square^{I^{(n)}} \chi I^{(n)}(X^{(n)}) dX^{(n)}, \text{ where } \chi I^{(n)}(X^{(n)}) = \begin{cases} 1 \text{ if } X^{(n)} \in \Omega^{I^{(n)}} \\ 0 \text{ otherwise} \end{cases} \quad (4)$$

A discretized version of the Lippmann-Schwinger equations can be written for each cluster as follows,

$$\Delta \boldsymbol{\varepsilon}^I(\mathbf{X}) = \Delta \tilde{\boldsymbol{\varepsilon}}^{(n)} - \sum_{J=1}^{k} \boldsymbol{D}^{I^{(n)}J^{(n)}} : \left[\Delta \boldsymbol{\sigma}^{J^{(n)}} - \mathbf{C}^{0,(n)} : \Delta \boldsymbol{\varepsilon}^{J,(n)}\right], n = 1, \ldots, N \quad (5)$$

Where the interaction tensor $\boldsymbol{D}^{I^{(n)}J^{(n)}}$ is defined as

$$\boldsymbol{D}^{I^{(n)}J^{(n)}} = \left[\frac{1}{C^I|\Omega^{(n)}|} \int_\Omega \int_\Omega \chi^{I^{(n)}}(\mathbf{X}) \chi^{J^{(n)}}(\mathbf{X}') \boldsymbol{\Gamma}^{(n)}(\mathbf{X},\mathbf{X}') d\mathbf{X}' d\mathbf{X}\right] \quad (6)$$

Once the interaction tensors among the clusters are calculated, the results are stored in an offline database. This precomputed offline interaction tensor database is later used to solve the cluster-based discretized Lippmann-Schwinger equation for any arbitrary loading conditions. This database can be reused for the part-scale calculation in an FE-MCA scheme where the part scale is modeled using a finite element solver considering different boundary conditions.

Table 1: Materials properties used for the micro-scale modeling[33]

|  | Matrix (PC) | Void | Fiber (SCF) |
|---|---|---|---|
| Elastic modulus (MPa) | 990 | 0.001 | $E_{11}$=235000, $E_{22} = E_{33}$ =14000, $G_{12} = G_{13} = 27000$, $G_{23} = 6500$ |
| Poisson's ratio | 0.33 | 0.33 | $\nu_{12} = \nu_{13} = 0.22, \nu_{23}$=0.3 |
| Hardening properties [$\varepsilon^p, \sigma$ (MPa)] | [0, 32.88]; [0.003, 34.30]; [0.009, 35.64] | 0 | - |

Figure 2. shows the original high-fidelity microstructure for the micro- and mesoscale of the problem. The clustering of the micro-scale microstructures is shown where the domain has three materials phases, namely the short carbon fiber (16 clusters), the polycarbonate matrix (16 clusters), and the voids (16 clusters). Each material's phase has been clustered separately. The materials properties considered for different materials phases are given in Table 1 where the SCF and voids are considered as an elastic materials phase and the matrix follows J2 plasticity calibrated against the following literature data [33]. For the mesoscale, the bead structures materials law is computed by the micro-scale homogenized response, and the mesoscale voids are modeled as an elastic phase with negligible stiffness. The mesoscale microstructure is decomposed with 32 clusters for the bead and a single cluster for the void phase.

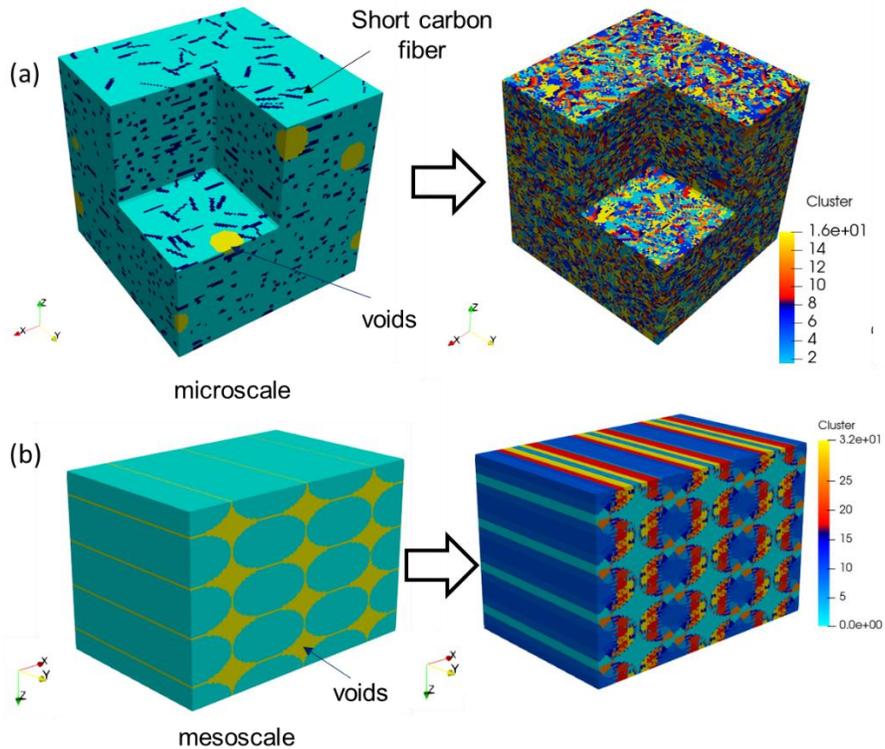

Fig. 2. Sample MVEs reconstruction for (a) microscale, (b) mesoscale microstructure, and ROM development through clustering

## 3. Results and discussions

In this section, the microscale intrabead void's interaction with the mesoscale interbead voids and how they influence the mechanical response of the parts for different mesoscale layer orientation designs are discussed with results consecutively.

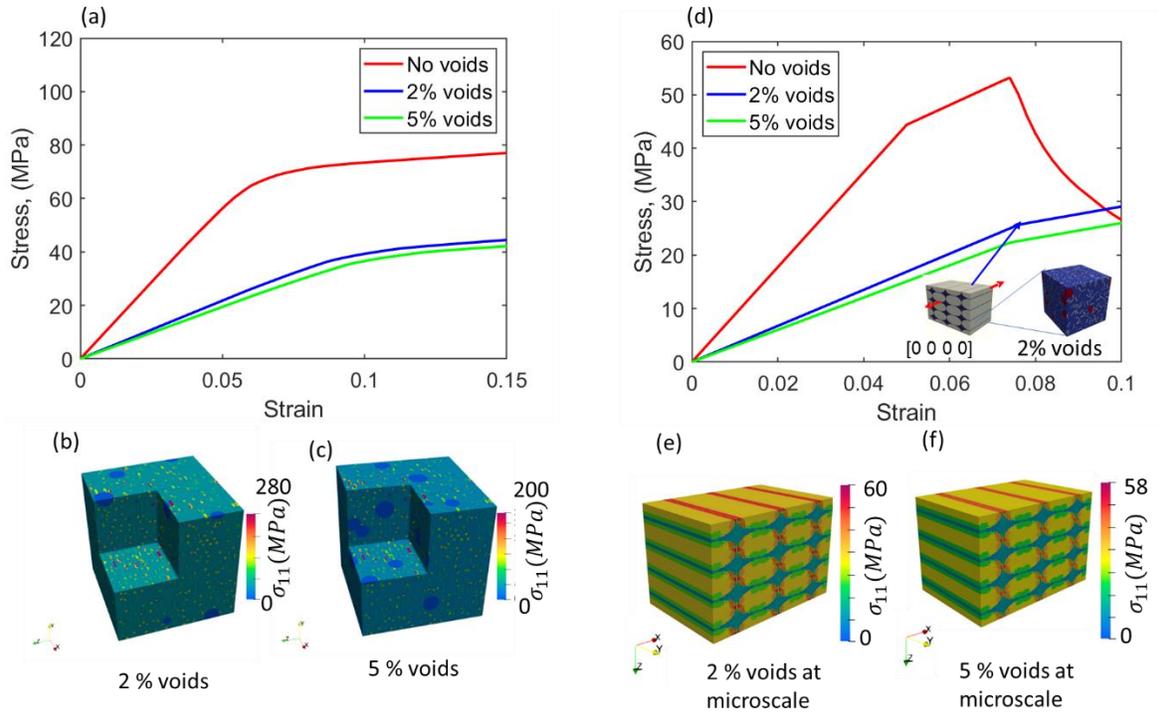

Fig. 3. (a) Micro-scale MVEs stress-strain response with and without defects and stress field at strain 0.1 for (b) 2% void, (c) 5% void. (d) Meso-scale MVEs stress-strain response with and without defects at micro-scale microstructure. Stress field at strain 0.05 for (e) 2% void, (f) 5% void.

### 3.1. Effect of defects on the microstructure

From the printed samples, we observed that voids are present in a single filament before and after the printing (Fig. 1d, 1f). These voids are generated due to trapped gas in the materials. First, the effect of the microscale voids on the micro-scale homogenized response is shown in Fig. 3a. The microstructure of the microscale is considered to have no defects, 2%, and 5% voids. Introducing 2% voids in the microstructure reduces the stiffness of the materials up to 62%. The localized response of the microscale microstructures is shown in Fig. 3b. The SCF introduces high stress as their stiffness is quite higher than the matrix. The homogenized response of the microscale is then passed to mesoscale microstructures with [0 0 0 0] bead orientation to understand the effect of micro-scale defects effect on the mesoscale response. As shown in Fig. 3c, the response of the mesoscale material further degrades due to the interaction of such meso- and micro-scale voids. The localized response of the mesoscale microstructure is shown in Fig. 3d.

## 3.2. Effect of mesoscale layer orientation

As the FFF process prints the composite layer by layer, the layer orientations influence the mechanical properties. The layer orientation at the mesoscale is varied for three different angles 0, 45, and 90 degrees. Four layers of the printed sample are considered for the mesoscale MVE, and tensile simulations are performed to obtain the stress-strain response. Fig. 4 shows that the [0 0 0 0] orientation of the bead shows better stiffness and tensile strength compared to the [45 45 45 45] and [90 90 90 90] orientation of the mesoscale beads. When the microscale is considered with defects, the mesoscale response again significantly degraded for all the different layer orientations.

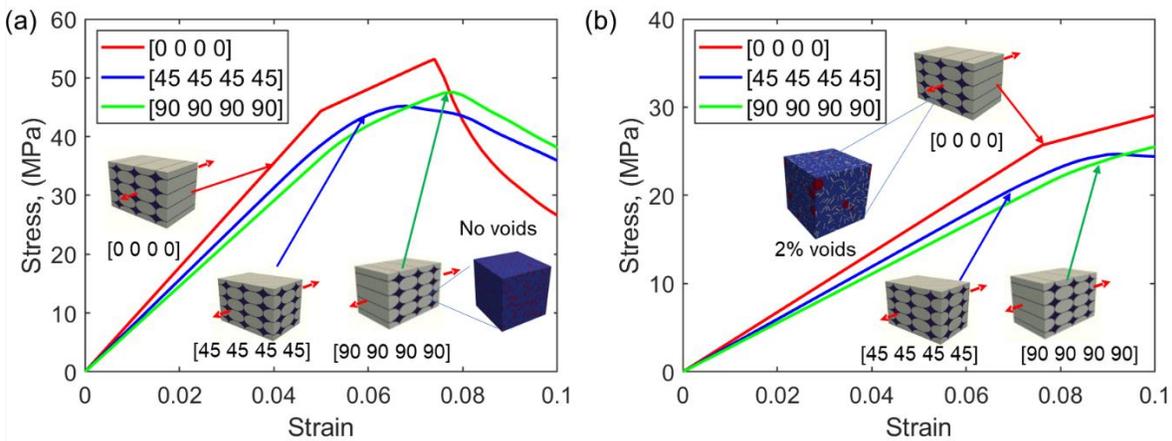

Fig. 4. Stress-strain response of mesoscale MVEs for different layer orientations with (a) no defects, and (b) 2% voids at micro-scale. The loading directions for MVEs and the micro-scale microstructure are also shown.

## 4. Summary

The as-built defects in PC/SCF composites fabricated using the FFF process have been studied at meso- and micro-scale by developing a mechanistic ROM. The defects at the meso-and micro-scale interact together and degrade the mechanical properties of the printed composites. Layer orientation of the printing is an important design parameter and loading applied along the bead printing direction shows better mechanical performance. Considering the high-dimensional design space, ROM provides a feasible way to study the multiscale design space of additively manufactured composites in a reasonable time and accuracy. Combining the multiscale materials design aspects with the FFF process, as shown in this work, provides a new avenue for the design and discovery of material systems

**Acknowledgments**

S.M. and W. K. L. thankfully acknowledge the support provided by AFOSR (FA9550-18-1-0381) and NSF (CMMI-1934367). A.B. is thankful to the University College Dublin's School of Mechanical and Materials Engineering Faculty Start-Up Grant (A298232/013813). ZI is thankful to Bowling Green State University Dept. of Engineering Technologies Faculty Start-Up Grant 33900032: SU Islam.